\documentclass[12pt,authoryear]{elsarticle}
 \pdfoutput=1
\usepackage{hyperref}
\usepackage{graphicx} 
\journal{Icarus}

\begin{document}

\begin{frontmatter}

\title{First experimental data of sulphur ions sputtering water ice}

\author[label1]{A. Galli\fnref{myfootnote}} \author[label1]{A. Vorburger}\author[label1]{P. Wurz}\author[label1]{R. Cerubini}\author[label1]{M. Tulej}

\fntext[myfootnote]{Corresponding author, E-mail address: \textit{andre.galli@space.unibe.ch}}

\address[label1]{Physikalisches Institut, University of Bern, Bern, Switzerland}

\begin{abstract}
This paper presents the first experimental sputtering yields for sulphur ions
with energies between 10 keV and 140 keV irradiating water
ice films on a microbalance. The measured sputtering yields exceed theoretical predictions based on other ion species by a 
factor of 2 to 3 for most energies.
Moreover, the sputtering yield of SF$^{+}$ molecules is compared to the yield of atomic species S$^{+}$ and F$^{+}$. 
As found for atomic versus molecular oxygen, the sputtering yield caused by molecules is two times higher than expected.
Finally, the implications of the enhanced sulphur sputtering yield 
for Europa's atmosphere are discussed.   
\end{abstract}


\end{frontmatter}

%\linenumbers

\section{Introduction}\label{sec:introduction}

When water ice is irradiated with energetic ions, the energy transferred from the impactor to the ice
may eject particles from the surface.
This process, termed sputtering, has been studied for several ion species and energy ranges
 under laboratory conditions over the past decades (e.g., \citet{har84b} and accompanying papers, \citet{bar03}, \citet{far05},
\citet{fam08}, \citet{joh09}, \citet{cas13}, \citet{mun16}, and \citet{gal17}).
Such results are needed to check theoretical models of sputtering \citep{sig69,joh89,cas05} and
relate them to observations \citep{hal95,han05,rot16} and models 
\citep{she05,smy06,pla12,vor18} of sputter-induced atmospheres of icy moons such as Europa.

Sputtering may proceed in a straight-forward manner, i.e., the ion directly knocking out one or several water molecules,
or it may be a two-stage process with the irradiation first causing chemical reactions inside the ice (so-called
radiolysis \citep{har84a,joh04,cas10}) and subsequently releasing the radiolysis products from the surface.
The sputtering yield denotes in both cases the number of water molecules or equivalents (if H$_{2}$O reacted to 
H$_{2}$ and O$_{2}$, for instance) leaving the ice per impacting particles. Knowing this yield and the chemical and energetic composition 
of the ejecta over a wide range of parameters is important to understand any ice-covered celestial body.
The astrophysical application we are most interested here is Europa, one of the icy moons of Jupiter.
For these bodies, the sputtering yields and the plasma environment determine the density and composition of their atmospheres
(see \citet{joh04} for a review).

For this study, we sputtered thin water ice films with sulphur ions and sulphur-bearing molecules from a microbalance.
This is the most common experimental method used so far
(see for example \citet{bar03,teo05,fam08,mun16,gal17}). However, to our knowledge,
sulphur ions were never used before in such experiments as sulphur is chemically reactive and can corrode surfaces
in vacuum chambers. Argon was the species closest in mass to sulphur so far, for which experimental results exist (see compilations by
\citet{joh09,cas13}). Knowing the sputtering yield of S$^{+}$ on water ice is highly relevant for Europa's atmosphere:
S$^{+}$ ions are one of the three most frequent ion species in the plasma
environment around Europa \citep{par02} because of its volcanically active neighbour Io.  

After a recapitulation of the theory of ice sputtering (Section \ref{sec:theory}), we describe the
experiment set-up in Section \ref{sec:experiment}.
We then present the sputtering results for S$^{+}$, F$^{+}$, and SF$^{+}$ molecules (Section \ref{sec:results}).
The paper is completed with a discussion on the implications of these results for Europa
and for future laboratory work (Section \ref{sec:implications}), followed by the conclusions (Section \ref{sec:conclusions}).      

\section{Theory}\label{sec:theory}

We will compare our new experimental results for S$^{+}$ and F$^{+}$ to two widely used semi-empirical formulae for ion sputtering yield derived by \citet{fam08} 
and \citet{joh09}. These formulae are based on previous experiments with other ion species irradiating dense water ice films.
For ion energies below 10 keV, the sputtering yield can be described 
by a cascade of elastic collisions, whereas the so-called electronic sputtering dominates at higher energies.
Equation~\ref{eq:sputteryield_fama} by \citet{fam08} is more accurate for lower energies;
for energies above 100 keV, Equation~\ref{eq:sputteryield_johnson} by \citet{joh09} offers a better fit to experimental data 
for H$^{+}$, He$^{+}$, N$^{+}$, O$^{+}$, Ne$^{+}$, and Ar$^{+}$ ion beams \citep{cas13}:

\vspace{-5mm}

\begin{equation}
Y(E,m_1,Z_1,\theta,T) = \frac{1}{U_i}\left( \frac{3}{4\pi^2C_0}\alpha S_n+\eta S^{2}_{e}\right)
    \left(1+q_i\exp\left(-\frac{E_a}{k_{B}T}\right)\right) \cos^{-f}(\theta)
\label{eq:sputteryield_fama}
\end{equation}

\noindent Equation \ref{eq:sputteryield_fama} quantifies the sputtering yield as a sum of nuclear and electronic
sputtering, described by the nuclear stopping cross section $S_n(E,m_1,Z_1)$ $=dE_n/(Ndx)$ and the electronic stopping cross section 
$S_e(E,m_1,Z_1)$. 
The yield in Eq.~\ref{eq:sputteryield_fama} depends on energy $E$, mass of impactor $m_1$, 
atomic number of impactor $Z_1$, the incidence angle $\theta$ relative to the surface normal, and surface temperature $T$. 
The temperature-dependent term with the activation energy $E_{a}$ \citep{rei84} becomes dominant above $T = 120$ K and is due
to radiolysis and subsequent release of H$_{2}$ and O$_{2}$ \citep{joh04, fam08, teo09}. This 
contribution makes up only 10\% of $Y$ at 90 K (Equation \ref{eq:sputteryield_fama}); it rises to 20\% at 100 K once 
the ice has been saturated with $\sim10^{15}$ ions cm$^{-2}$ \citep{teo05}. 
For $U_i$, the sublimation energy of water (0.45 eV) is assumed.
The effective cross-section for low energy recoils, $C_0=1.3$ \AA$^{2}$,
the activation energy, $E_a =0.06\pm0.01$ eV, and
$q_i=220$ are constants \citep{fam08}. The parameter describing the
angular dependence calculates to $f=1.75$ for S$^{+}$.

By comparison, \citet{joh09} propose the following empirical formula for the sputtering yield (the angular and temperature
dependence are identical to Equation~\ref{eq:sputteryield_fama}):

\vspace{-5mm}

\begin{equation}
Y(E,m_1,Z_1,\theta,T) = 1/(1/Y_{low}+1/Y_{high})
    \left(1+q_i\exp\left(-\frac{E_a}{k_{B}T}\right)\right) \cos^{-f}(\theta)
\label{eq:sputteryield_johnson}
\end{equation}

\noindent whereby $Y_{low}$ and $Y_{low}$ stand for:

\vspace{-5mm}

\begin{equation}
Y_{i} = Z_{1}^{2.8} C_{1}\left(\frac{v}{2.19\times 10^{6}}Z_{1}^{-1/3}\right)^{C_{2}}, \textup{ with ion velocity }v = \sqrt{2E/m_{1}}
\end{equation}

\noindent The fit parameters are $C_1= 4.2$ and $C_2 = 2.16$ for $Y_{low}$ and $C_1= 11.22$ and $C_2 = -2.24$ for $Y_{high}$.

\section{Experiment set-up}\label{sec:experiment}

The same microbalance set-up was used as in our previous ice film experiments \citep{gal17}:
We background deposited de-ionized water vapour via a needle valve and a capillary
onto the cooled surface of a microbalance in a vacuum chamber. 
The sensitivity of the microbalance was $1.61 \times 10^9$ Hz g$^{-1}$ according to 
calibration in 2014 performed by the manufacturer (gold-coated 15 MHz quartz crystal, manufacturer: QCM Research).
The surface of the microbalance was 45$^{\circ}$ or 60$^{\circ}$ tilted with respect to the incoming ion beam. 
Under these conditions and
temperatures around 90~K, most of the deposited ice will remain amorphous throughout 
the experiments and the porosity will
vary between few \% \citep{fam08} and 25\% \citep{mit17}. If the bulk density is
0.9 g cm$^{-3}$, one monolayer of H$_{2}$O on the microbalance corresponds to a frequency shift of 14 Hz.
The H$_{2}$O partial pressure in the chamber ranged from $3\times10^{-8}$ to $4\times10^{-7}$ mbar during vapour deposition;
the ice film accretion rate increased linearly with that partial pressure from 0.2 to 5 Hz s$^{-1}$. 
Within this range, the deposition rate did not notably affect the measured sputtering yields. 
For irradiation experiments, the ice film thickness ranged from 40 to 200 nm and 
the residual water pressure in the vacuum chamber was only on the order of $10^{-9}$ mbar.
%For reliable measurements, the ice film should be thicker than the ion penetration depth 
%at a given energy \citep{gal17}.

To create an ion beam, we ionized SF$_{6}$ gas and accelerated the ion species with an electron-cyclotron-resonance ion source \citep{mar01}. 
The ion source produced many different species from the parent molecule SF$_{6}$, 
but only S$^{+}$, F$^{+}$, S$^{2+}$ and SF$^{+}$ turned out to have a sufficiently high beam current to create
a detectable sputtering signal when the ion beam was directed at the water ice film on the microbalance.
The beam currents reached 0.1 to 1.0 nA, which corresponded, at a beam diameter of 0.3 cm, to
$(0.9\dots9) \times 10^{10}$ ions cm$^{-2}$ s$^{-1}$. To interpret the results for the 
SF$_{6}$ fragments and to verify the microbalance sensitivity, 
we also irradiated the microbalance with O$^{+}$ ions whose sputtering yield is well known 
from previous studies \citep{shi95,bar03}.

\section{Results}\label{sec:results}

\subsection{Accuracy of results}\label{sec:reproducibility}

Before we discuss the results for the hitherto unknown sputtering yields of S$^{+}$, F$^{+}$, and SF$^{+}$, 
let us first assess the general accuracy of our experiments using oxygen as a reference. 
Oxygen sputtering yields have been measured numerous times by us and other research groups whereby Equation \ref{eq:sputteryield_fama} fits most previous laboratory experiments
within 30\% relative uncertainty \citep{fam08}.
We therefore collected all O$^{+}$ yield results measured at an impact angle of $45^{\circ}$ 
and 10, 30, and 50 keV energy over the last 1.5 years in our facility. 
This data set was accumulated during five different measurement series separated by several weeks or months. The first part of this data set covering the year 2016
was presented in \citet{gal17}; here we added the measurements from 2017.
We normalized all data (obtained at temperatures between 89 and 101 K) to the same temperature $T = 90$ K assuming 
the temperature dependence in Equations \ref{eq:sputteryield_fama} and \ref{eq:sputteryield_johnson}.
Apart from temperatures also vacuum pressure, vapour deposition rate, ice film thickness, and irradiation duration varied.
Moreover, we used two different microbalances of the same type (see Section \ref{sec:experiment}).

The average sputtering yields derived from this comprehensive data set compared to the 
$Y_{th}$ predicted from Equation \ref{eq:sputteryield_fama} (with $T = 90$ K) as follows:
$Y = 44\pm13$ (14 data points) vs. $Y_{th}=27$ at 10 keV, $Y = 73^{+14}_{-25}$ (7 data points) vs. $Y_{th}=62$ at 30 keV, 
and $Y=111^{+15}_{-33}$ (5 data points) vs. $Y_{th}=105$ at 50 keV. 
The experimental error bars denote the ranges between average and most extreme
positive and negative outlier.
 
These results are important in two respects: 
First, the experimental values are reproducible within 30\% or better on the long run.
During one measurement series, the scatter usually was on the order of 10\% \citep{gal17}.
Second, the oxygen results agree with Equation \ref{eq:sputteryield_fama}, which reproduces previous results 
from other groups within 30\% \citep{fam08}.  
We will therefore attribute an uncertainty of 30\% to our experimental yields for other ion species in the following section.

\subsection{Results for S$^{+}$, F$^{+}$, and SF$^{+}$}

\begin{figure}
\begin{center}
\includegraphics[clip,width=1.0\textwidth]{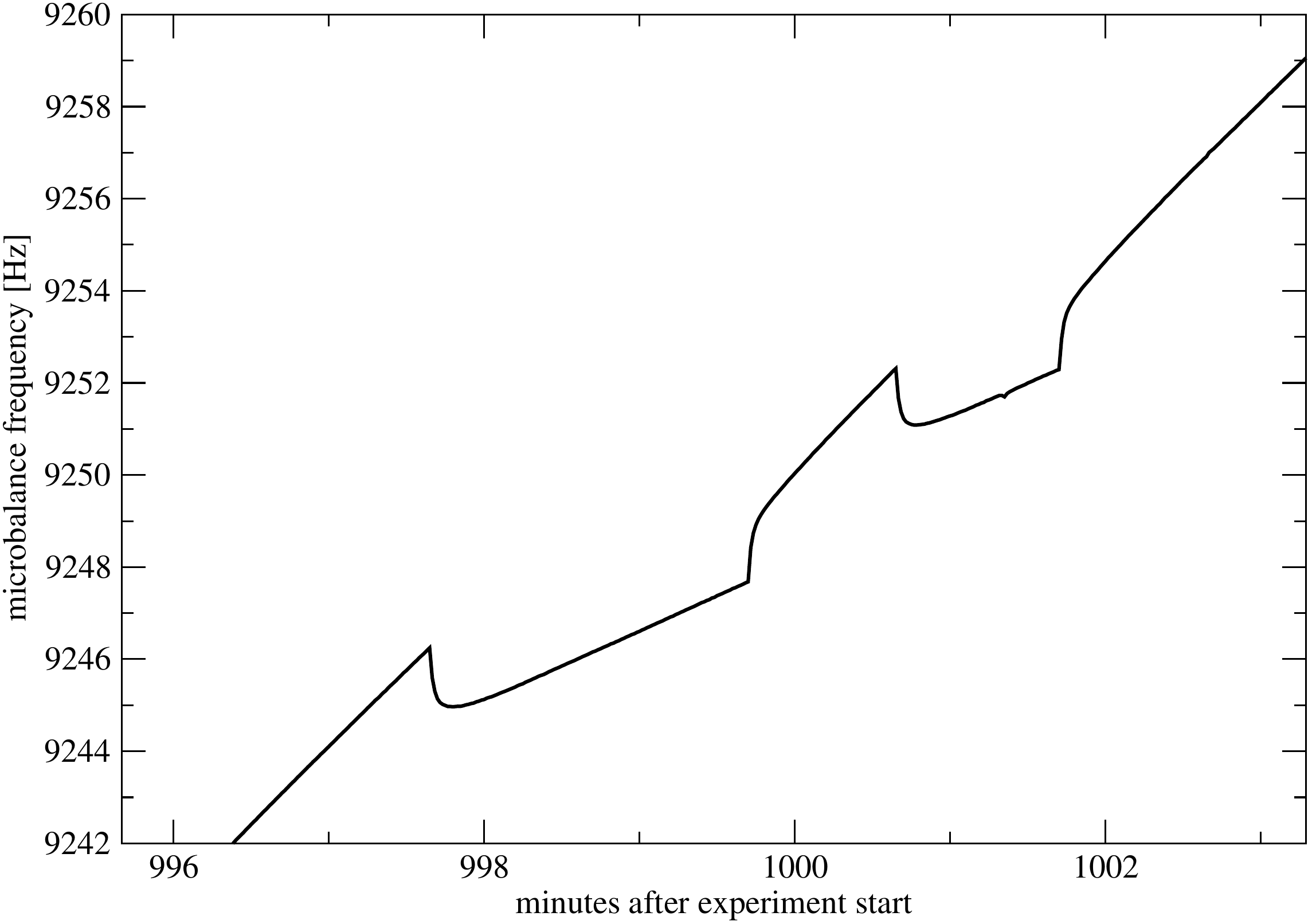}
\end{center}
\caption{Frequency response of the microbalance to a 50 keV S$^{+}$ ion beam for 2 and 1 min
of irradiation. The ice accretion rate on the microbalance flattens because of the sputtering losses.
The momentum of the ion beam also causes an immediate frequency drop (at minute 998 and 1001)
when the ion beam hits the microbalance and a frequency rise when the ion beam is shut off (at minute 1000 and 1002).}\label{fig:50keVsulfurexample} 
\end{figure}

105 individual irradiations with S$^{+}$, SF$^{+}$, S$^{2+}$, and F$^{+}$ ions
hitting the ice film were performed during two different measurement series of six days in total. 
11 of the S$^{+}$ irradiations took place the same day when also O$^{+}$ sputtering yields were measured for 
cross-calibration (see Section~\ref{sec:reproducibility}). 
The median temperature of the water ice film during the measurements was 90 K with extremes of 89 K and 101 K.
For evaluation, individual sputtering yields obtained at a temperature other than 90 K were normalized to $T = 90$ K based on 
Equations~\ref{eq:sputteryield_fama} and \ref{eq:sputteryield_johnson}. At the given temperature range this corresponds to a modification by a factor of 1.1
at most.
An individual irradiation lasted between 1 and 30 minutes; Fig.~\ref{fig:50keVsulfurexample} 
shows as an example the frequency response of the microbalance (in Hz versus minutes) to a 50 keV S$^{+}$ ion 
beam for 2 and 1 minutes of irradiation. With these measurements, we derived the sputtering yield 
from the difference of the accretion rate during irradiation compared to the rate before and after irradiation when the ion beam was off.
 
After depositing an ice film, we irradiated it several times to reach fluences of $1.8\times10^{14}$ 
ions cm$^{-2}$ at most. We then either deposited a fresh ice film onto the irradiated ice film (repeating this step one or two times)
or we desorbed all ice layers before depositing a pristine ice film by heating and cooling the microbalance. 
The derived sputtering yields did not change notably for the two different cases.
Due to the build-up of O$_{2}$ in the ice film, irradiation at higher ion fluence results in higher sputtering yields \citep{teo05}.
At 90 K ice temperature, however, this contribution is expected to enhance the total sputtering yield by only 10\% for saturated ice. 
Moreover, the highest
ion fluence in our measurements corresponds to only 20\% of the saturation fluence (see Section \ref{sec:theory}).
We quantified the fluence effect on sputtering yields in two ways: During 30 minutes of continuous irradiation with a 30 keV S$^{+}$ beam, 
the sputtering yields varied by less than 10\%.  
We also compared the sputtering yields from subsequent irradiations
at identical parameters. For 43 different pairs of irradiations, the second attempt resulted in a $1.06\pm0.09$
higher sputtering yield than the first attempt. In summary, varying ion fluences affected the sputtering yields
presented here by 10\% at most. This agrees with the estimate for O$^{+}$ and Ar$^{+}$ irradiation \citep{gal17}. Hence we did 
not discriminate data according to fluence and organized the data points solely according to different ion
species, impact angle ($45^{\circ}$ and $30^{\circ}$ relative to surface normal), and energy (10 keV to 140 keV).
This approach resulted in 22 different settings to be distinguished.

The effect of ice film thickness is taken into account the same way as in \citet{gal17}: 
We relied on SRIM calculations \citep{zie85,zie08} to estimate the average penetration depth of ions. 
In the results (Table~\ref{tab:yields_ions}), 
we included only those sputtering yields measured
with an ice film thickness deeper than the penetration depth: sputtering yields derived  
from irradiations at an insufficient film thickness result in a notably larger scatter (refer to 
Fig. 1 in \citet{gal17}).
The only exception was the 
measurement with 140 keV S$^{2+}$. Here, the ice film was only 0.4 times the expected penetration
depth. 
The study by \citet{gal17} about Ar$^{+}$ and O$^{+}$ sputtering showed that the measured sputtering yield 
for irradiations with that ratio of ice film thickness to penetration depth was 0.9 times the value measured 
for thicker ice films. 
The yield and its uncertainty for 140 keV S$^{2+}$ therefore might be larger than stated in Table \ref{tab:yields_ions}.
For the SF$^{+}$ molecule, we assumed that it fragments upon impact and then used the kinetic energy of the S fragment 
(44 keV for the case of a 70 keV SF$^{+}$ molecule, e.g.) to estimate the penetration depth.
% More precisely: Assume that 19/30 of the energy are transferred to the S+ fragment and then interpolate
% the penetration depths from the calculated depths for S+:
% This leads to: 70 keV SF+ -> 44 keV S+ which would imply d = 70+-27 nm (or 26 keV F+, which implies 75+-30)
%                50 keV SF+ -> 32 keV S+ -> 58 +-20
%                30 keV SF+ -> 19 keV S+ -> 34 +-14
%                20 keV SF+ -> 13 keV S+ -> 24 +-10
%                10 keV SF+ -> 6.3 keV S+ -> 14 +- 6
% To evaluate sputtering yield we were conservative and relied only on those data points obtained at thickness > penetration depth
% of sulfur with the full energy (e.g. 70 keV instead of 44 keV etc.)

Regarding F$^{+}$, only for one setting with 30 keV did we achieve a thickness $\approx$ penetration depth.
%the penetration depth for 70 keV F+ at 45 deg impact angle would be $196\pm67$ nm and $84\pm34$ for 30 keV. 
The values in Table \ref{tab:yields_ions} are the averages over the individual irradiations, the error bar is always 30\%
as derived in Section~\ref{sec:reproducibility}.

The resulting 22 values of $Y$ for the different data sets are presented in Table \ref{tab:yields_ions}.
The impact angle $\theta$ is the angle relative to the surface normal as in Equation \ref{eq:sputteryield_fama}.
The energy in Table \ref{tab:yields_ions} is the total kinetic energy per ion, corrected for ion charge state.
Based on previous experiments \citep{mun16,gal17}, we assumed that charge state has no effect on sputtering yield
and used Eq. \ref{eq:sputteryield_fama} to predict the sputtering yield for 140 keV S$^{2+}$ accordingly.
The nuclear and electronic stopping cross-sections, $S_n$ and $S_e$, were added for S$^{+}$ 
and F$^{+}$ in units of eV \AA$^{2}$ following the approach in the Appendix of \citet{fam08}.

\begin{table}
\begin{center}
\caption{Average sputtering yields $Y$ for all different experiment settings: ion species, energy $E$, 
and impact angle $\theta$. The theoretical values $Y_{th}$ were calculated
with the semi-empirical formula in Eq. \ref{eq:sputteryield_fama}, 
assuming charge state does not affect sputtering yields. Data and theoretical values were scaled to 
the same temperature $T = 90$ K with Equations~\ref{eq:sputteryield_fama} and \ref{eq:sputteryield_johnson}.
The yield for S$^{2+}$ with the asterisk
might be 10\% larger because of the insufficient ice film thickness at these high energies.
No theoretical predictions for molecules such as SF exists yet.
The average penetration depth and its uncertainty are listed as $d$ and 
$\sigma_d$, followed by the nuclear and electronic stopping cross-sections $S_n$ and $S_e$ 
from Eq.~\ref{eq:sputteryield_fama}.}
\label{tab:yields_ions}
\begin{small}
\begin{tabular}{lrrrrrrrrr}
\hline
Ion & $E$ (keV) & $\theta$ ($^{\circ}$) & $Y$ & $\sigma_{Y}$ & $Y_{th}$ & $d$ (nm) & $\sigma_d$ (nm) & $S_n$ (eV \AA$^{2}$) & $S_e$ (eV \AA$^{2}$)\\
% New evaluation on November 22:
% 1) New shots from November 2017 added (11 additional S+ data points)
% 2) Scaled all Yields to 90K because of temperature range from 101 K to 89 K (it is a 10% effect at most for 100 K)
% 3) Scaled the different measurement series to have the same frequency lapse response of TQCM to an ion beam of known current
%    as in 2016. With this approach, the yields for O+ are consistent within a range of 30% over 2 years of data (and
%    six different measurement campaigns) and they agree at 30 keV and 50 keV with Fama et al. 2008 (calibrated to other research groups) 
% 4) By default, all yield uncertainties are 30% of the value. This is conservative but not too pessimistic given the 2 years
%    intercalibration studies and the comparison of S+ and O+ yields between different series and with Fama et al. 2008 predictions for O+
S$^{+}$	& 70	& 45	& 144	& 43& 52	& 104	& 39 & 27 & 17 \\ % Case 37 in database
S$^{+}$	& 50	& 45	& 105	& 32& 43	& 79	& 28 & 30 & 14 \\
S$^{+}$	& 30	& 45	& 94	& 28& 35	& 56	& 20 & 33 & 10 \\
S$^{+}$	& 20	& 45	& 78	& 23& 32	& 35	& 14 & 35 & 8 \\
S$^{+}$	& 10	& 45	& 46	& 14& 28 	& 22	& 9 & 35 & 5 \\
S$^{+}$	& 70	& 30	& 110	& 33& 36	& 129	& 41 & 27 & 17 \\
S$^{+}$	& 50	& 30	& 98	& 29& 30	& 96	& 30 & 30 & 14 \\
S$^{+}$	& 30	& 30	& 80	& 24& 24	& 60	& 20 & 33 & 10 \\
S$^{+}$	& 20	& 30	& 65	& 20& 22 	& 42	& 15 & 35 & 8 \\
S$^{+}$	& 10	& 30	& 43	& 13& 20 	& 24	& 9 & 35 & 5 \\
S$^{2+}$& 140	& 45	& *170	& 51& 94	& 213	& 67 & 21 & 26 \\
SF$^{+}$& 70	& 45	& 287	& 86& N/A	& 70	& 27 & N/A & N/A \\
SF$^{+}$& 50	& 45	& 265	& 80& N/A	& 58	& 20 & N/A & N/A \\
SF$^{+}$& 30	& 45	& 234	& 70& N/A	& 34	& 14 & N/A & N/A \\
SF$^{+}$& 20	& 45	& 157	& 47& N/A	& 24	& 10 & N/A & N/A \\
SF$^{+}$& 10	& 45	& 112	& 34& N/A	& 14	& 6 & N/A & N/A \\
SF$^{+}$& 70	& 30	& 229	& 69& N/A	& 87	& 28 & N/A & N/A \\
SF$^{+}$& 50	& 30	& 206	& 62& N/A	& 70	& 21 & N/A & N/A \\
SF$^{+}$& 30	& 30	& 186	& 56& N/A	& 36	& 14 & N/A & N/A \\
SF$^{+}$& 20	& 30	& 142	& 43& N/A	& 29	& 11 & N/A & N/A \\
SF$^{+}$& 10	& 30	& 102	& 31& N/A	& 17	& 7 & N/A & N/A\\
F$^{+}$	& 30	& 45	& 66	& 20& 53 	& 84	& 34 & 14 & 9\\
\end{tabular}
\end{small}
\end{center}
\end{table}

The sputtering yields of S$^{+}$ and SF$^{+}$ increase with roughly $\cos^{-1}(\theta)$
(see Equations \ref{eq:sputteryield_fama} and \ref{eq:sputteryield_johnson}) for an increase of impact angle $\theta$
from 30$^{\circ}$ to 45$^{\circ}$. 
This is a weaker dependence than the range of $f=1.3\dots1.8$ found by \citet{fam08} and \citet{vid05} for a wider range of angles and impactors.
Since we have only two different angular positions at our disposal, we cannot conclude if the angular dependence for
S$^{+}$ and SF$^{+}$ truly deviates from the hitherto established angular dependence for other sputtering species.  
%One possible explanation is that our ice film might have a rough, uneven surface similar to frost because of the remote vapour deposition technique. 
%In the following, we will concentrate on the results obtained at $45^{\circ}$ to compare our measured 
%yields with predictions \citep{fam08,joh09}
%at an intermediate angle between the two extremes of head-on collisions and grazing impacts \citep{kue98}.}
% For S+, 70 keV, we have a ratio of 1.25
% For 50 keV, r = 1.18
% For 30 keV, r = 1.25
% For 20 keV, r = 1.20
% For 10 keV, r = 1.19, thus energy-independent ratio of 1.21+-0.04 -> f in cos^-f calculates to f = 0.95+-0.15

\begin{figure}
\begin{center}
\includegraphics[clip,width=1.0\textwidth]{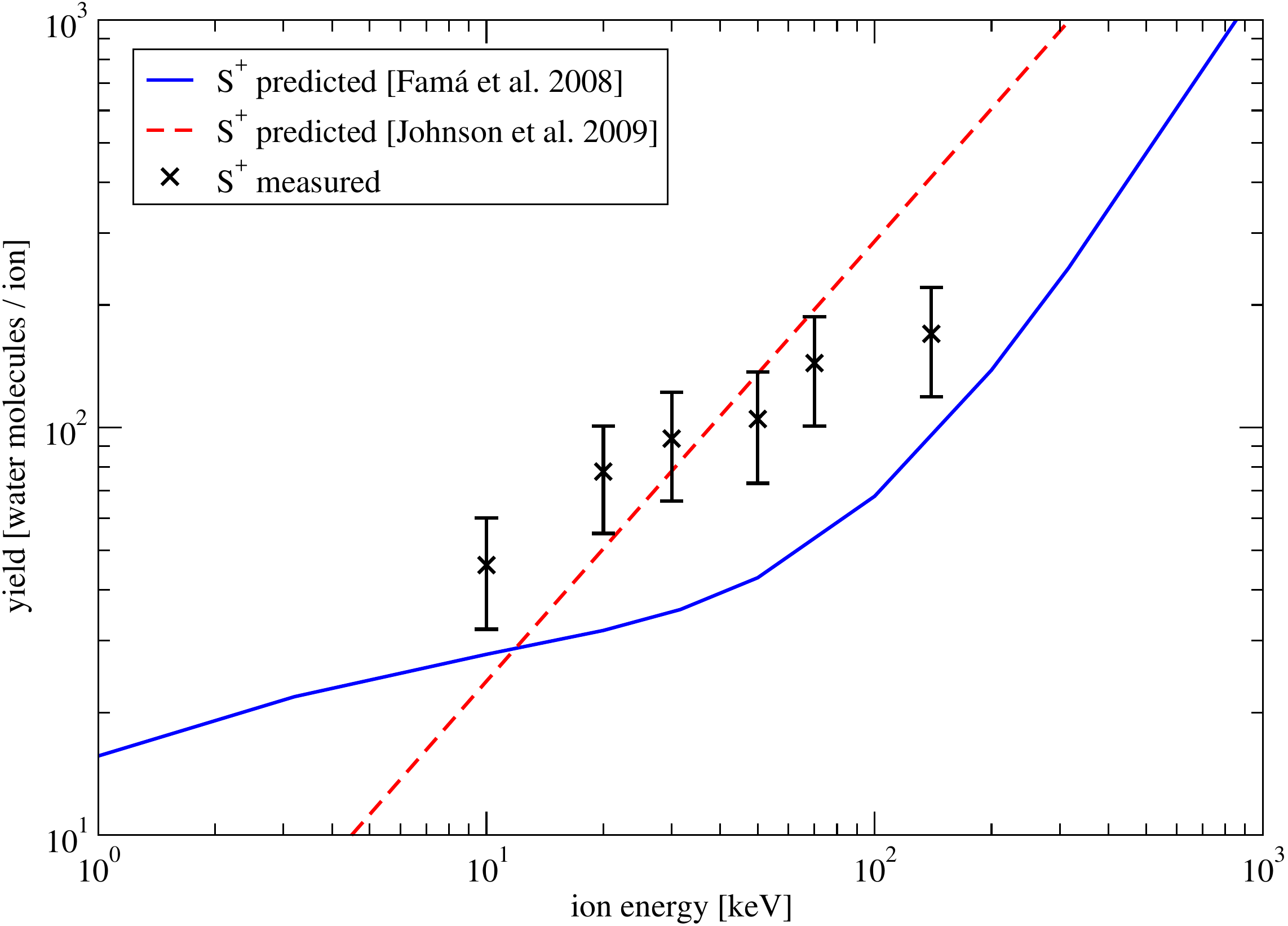}
\end{center}
\caption{S$^{+}$ sputtering yields versus energy for $45^{\circ}$ impact angle and $T= 90$ K. Data points (black symbols) are taken from the present study;
the blue curve shows the prediction by \citet{fam08} and the red dashed curve shows the prediction by \citet{joh09}, both of which
are based on previous ice sputtering experiments with other ion species.}\label{fig:fama}
\end{figure}

Figure~\ref{fig:fama} compares the experimental S$^{+}$ sputtering yields to the predictions of 
Eq.~\ref{eq:sputteryield_fama} \citep{fam08}
(blue curve) and Eq.~\ref{eq:sputteryield_johnson} \citep{joh09} (red dashed curve) for the impact angle of $45^{\circ}$.
%The values from Eq.~\ref{eq:sputteryield_fama} are identical to the $Y_{th}$ in Table~\ref{tab:yields_ions} for $\theta=45^{\circ}$.
The experimental sputtering yield exceeds
the predictions by \citet{fam08} by a factor of $2.9\pm0.5$ for all energies between 20 and 70 keV. The discrepancy
is less notable for 10 and 140 keV. 
By comparison, the formula by \citet{joh09} matches the data at 30 and 50 keV.
For lower energies where elastic collisions dominate, the predictions by \citet{joh09}
also underestimate the sputtering yield by a factor of two. The tendency of Equation~\ref{eq:sputteryield_johnson} to underestimate 
yields at low energies was already noted by the authors and by \citet{cas13}. More noteworthy 
is that the sputtering yield between 10 and 70 keV exceeds predictions from Equation~\ref{eq:sputteryield_fama}.
That equation matches experiments better than Equation~\ref{eq:sputteryield_johnson}
below 100 keV energy in the case of H$^{+}$, He$^{+}$, N$^{+}$, O$^{+}$, and Ar$^{+}$ \citep{cas13}.
The implications of these enhanced sulphur sputtering yields for Europa's atmosphere will be discussed in Section \ref{sec:implications}.

The one experiment result for F$^{+}$ at 30 keV agrees with predictions by \citet{fam08}.

%\begin{figure}
%\begin{center}
%\includegraphics[trim={1cm 2cm 3cm 2cm},clip,width=1.0\textwidth]{SFtoS_ratio.pdf}
%\end{center}
%\caption{Ratio of sputtering yields from SF$^{+}$ versus S$^{+}$ for identical energies and impact angles
%(see Table~\ref{tab:yields_ions}). The red dashed line is the average ratio of $2.25\pm0.23$.}\label{fig:sf_to_s_ratio}
%\end{figure}

The measured sputtering yields for SF$^{+}$ are much higher than those for S$^{+}$ and F$^{+}$ 
at the same energies and impact angles. A quantitative interpretation
of the SF$^{+}$ yields is difficult because they have never been measured before. Moreover,
there is, to our knowledge, no general theory predicting sputtering yields
from molecules irradiating solid surfaces. Comparing sputtering yields
from atomic and molecular oxygen in water ice \citep{gal17}, we noted that the 
measured sputtering yield for an oxygen molecule was roughly $4Y(E/2)$ for energies
between 10 and 50 keV. This is two times higher than the $2Y(E/2)$ one would predict 
under the assumption that the impact fragments the O$_{2}$ molecule and distributes 
the available kinetic energy equally unto the fragments.

In the present study, we find a similar excess for $Y($SF$^{+})$ compared to the sputtering
with atomic S$^{+}$ and F$^{+}$. We considered the SF$^{+}$ measurements at an impact angle of $45^{\circ}$ 
in Table~\ref{tab:yields_ions}
and assumed that the kinetic energy of the molecule is re-distributed to its fragments upon 
impact according to the mass ratio. This implies, e.g., that a 30 keV SF$^{+}$
is fragmented into an 11 keV F$^{+}$ and a 19 keV S$^{+}$. We then applied Equation~\ref{eq:sputteryield_fama},
multiplied by the observed enhancement of 1.25,
 to calculate $Y($F$^{+})$ and we interpolated $Y($S$^{+})$ from Figure~\ref{fig:fama} 
to the relevant energies. The sum of the thus calculated monoatomic yields $Y($S$^{+})$ + $Y($F$^{+})$ 
is 83, 110, 141, and 159 for molecule energies of 20, 30, 50, and 70 keV, respectively. 
The ratio of measured molecular yields ($Y($SF$^{+})$ in Table~\ref{tab:yields_ions}) 
to these predictions calculates to $1.93\pm0.14$. The physical reason for this enhancement is currently unknown. 

\section{Implications for Europa}\label{sec:implications}

\begin{figure}
\begin{center}
\includegraphics[clip,width=1.0\textwidth]{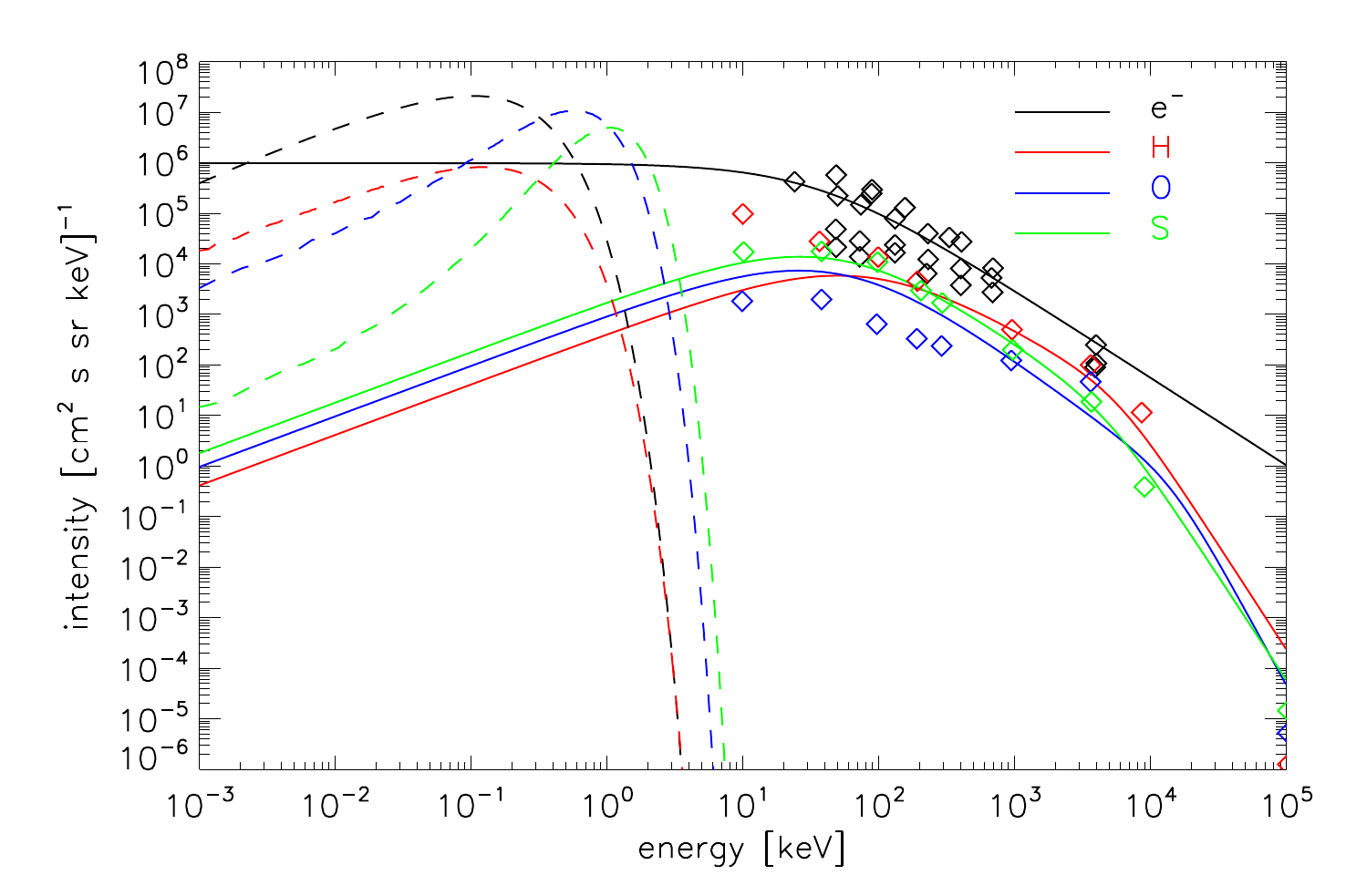}
\end{center}
\caption{Intensity spectra of the four major plasma species (black: electrons, red: H$^{+}$, blue: O$^{+}$, green: S$^{+}$)
at Europa's surface. The diamonds denote data from \citet{ip98} (ions) and \citet{par01} (electrons),
the solid lines are fits to the hot populations, the dashed lines are models of the cold populations that have not been
measured yet. Figure adapted from \citet{vor18}.}\label{fig:audrey}
\end{figure}

If sputtering yields of energetic sulphur ions irradiating water ice are higher than assumed so far, the
density of molecules in Europa's atmosphere predicted by models will increase. We estimate
this increase in this section.  
The main plasma constituents irradiating Europa's icy surface are 
electrons, H$^{+}$, O$^{+}$, and (from the nearby 
moon Io) S$^{+}$ \citep{ip98,coo01,par02}. All of these species form two separate distributions, a cold population
with modes typically around 1 keV, and a hot population with modes around 100 keV.
This is illustrated in Fig.~\ref{fig:audrey} from \citet{vor18}. 
Ion intensity measurements (red, blue, and green diamonds) 
were taken from \citet{ip98} whereas electron intensity measurements (black diamonds) were taken
from \citet{par01}. The cold plasma energy spectra were fitted with drifting
Maxwellian distributions (dashed lines) whereas the hot plasma energy spectra were
fitted with Kappa distributions (solid lines). Also shown, for comparison, is the
electron spectral shape as presented in \citet{par01} (dotted black lines).

According to the model by \citet{vor18}, cold ions, hot ions, cold electrons, and hot electrons all
sputter the same order of magnitude H$_{2}$ and O$_{2}$ into the atmosphere, leading to a surface particle density between
$10^{3}$ and $10^{4}$ cm$^{-3}$ for O$_{2}$ and H$_{2}$. In addition, the cold and hot ions also eject H$_{2}$O with a similar surface density. 
As a simple sensitivity test, we increased the sputtering yields for the hot sulphur population (between
a few keV and a few 100 keV) by a factor of 3 compared to assumptions so far. This enhancement is consistent with the new experimental
results. The sputtering yields for the cold sulphur population 
and for all other plasma species remained the same in the model.
This would have the following impacts on the model predictions by \citet{vor18}:
The total surface density (average temperature assumed to be 106 K) of the sputtered species at Europa increase from
$1.0$ to $1.4\times10^{4}$ cm$^{-3}$ (H$_{2}$O), $0.9$ to $1.1\times10^{4}$ cm$^{-3}$ (H$_{2}$), and from $1.6$ to $1.9\times10^{4}$ cm$^{-3}$ (O$_{2}$).
The atmosphere surface densities increase only by 30\% in total because sulphur is not the only sputtering agent at Europa's surface.
%  OLD = 6e3 + 4e3 = 1.0e4; NEW: 6e3 + 8e3 = 1.4e4
%OLD = 2e3 + 2e3 + 4e3 + 6e2 = 0.9e4; NEW = 2e3 + 4e3 + 4e3 + 6e2 = 1.1e4
%OLD: 4e3 + 3e3 + 7.5e3 + 1.5e3 = 1.6e4; NEW: 4e3 + 6e3 + 7.5e3 + 1.5e3 = 1.9e4

This discussion only considered the effect of sulphur sputtering for the ejected particles and thus for Europa's atmosphere.
The effects of sulphur ion irradiation on the chemical and physical properties of Europa's surface will be studied
in future tests with chemically more complex ice films and with thick ice samples \citep{gal16}. 

\section{Conclusions}\label{sec:conclusions}

The first experimental results for sulphur ions sputtering of water ice show
that the sputtering yield between 20 and 70 keV 
is two to three times higher than predicted by extrapolations with other ion species \citep{fam08,joh09,cas13}.
This has some implications for the Jovian system, in particular for the atmosphere of Europa. 
The enhanced sputtering yields reported here imply that the pressure of sputtered
O$_{2}$, H$_{2}$, and H$_{2}$O at Europa's surface exceed previous model predictions by roughly 30\%, all else being equal. 
Comparing the sputtering yield from SF$^{+}$ to S$^{+}$ and F$^{+}$ between 20 and 70 keV, 
we find that sputtering induced by molecules 
is two times more efficient than expected from the sputtering yield of monoatomic species.
This is the same finding as in previous sputtering experiments by \citet{gal17} for atomic versus molecular oxygen.  

\section*{Acknowledgements}
The work in this paper has been partially performed in the context of
the activities of the ISSI International Team Nr. 322, www.issibern.
ch/teams/exospherejuice/. We also would like to thank G. Bodmer,
J. Gonseth, and A. Etter for their relentless support of the scientific
work at the MEFISTO facility.

%\section*{References}


\begin{thebibliography}{}

\bibitem[Baragiola et al.(2003)]{bar03}Baragiola, R.A. Vidal, R.A., Svendsen, W., Schou, J.,
Shi, M., Bahr, D.A., Atteberry, C.L., 2003. Sputtering of water ice. 
Nuclear Instruments and Methods in Physics Research B 209, 294--303.

\bibitem[Cassidy and Johnson(2005)]{cas05}Cassidy, T.A., Johnson R.E., 2005. 
Monte Carlo model of sputtering and other ejection processes within a regolith. Icarus 176, 499--507.

\bibitem[Cassidy et al.(2010)]{cas10}Cassidy, T., Coll, P., Raulin, F., Carlson, R.W., Johnson, R.E., Loeffler, 
M.J., Hand, K.P., Baragiola, R.A., 2010. Radiolysis and Photolysis of Icy Satellite Surfaces: 
Experiments and Theory. Space Science Reviews. doi:10.1007/s11214-009-9625-3.

\bibitem[Cassidy et al.(2013)]{cas13}Cassidy, T.A., Paranicas, C.P., Shirley, 
J.H., DaltonIII, J.B., Teolis, B.D., Johnson, R.E., Kamp, L., Hendrix, A.R., 2013.
Magnetospheric ion sputtering and water ice grain size at Europa. Planetary and Space Science 77, 64--73. 

\bibitem[Cooper et al.(2001)]{coo01}Cooper, J.F., Johnson, R.E., Mauk, B.H.,
Garrett, H.B., Gehrels, N., 2001.
Energetic Ion and Electron Irradiation of the Icy Galilean Satellites. Icarus 149, 133--159.

\bibitem[Fam\'{a} et al.(2008)]{fam08}Fam\'{a}, M., Shi, J., Baragiola, R.A., 2008. Sputtering of ice by low-energy ions. 
Surface Science 602, 156.

\bibitem[Farenzena et al.(2005)]{far05}Farenzena, L.S., Iza, P., Martinez, R., Fernandez-Lima, F.A.,
Seperuelo Duarte, E., Faraudo, G.S., Ponciano, C.R., da Silveira E.F., 
C.R., Homem, M.G.P., Naves de Brito, A., Wien, K., 2005. Electronic sputtering analysis of astrophysical ices. 
Earth, Moon, and Planets 97, 311--329.

\bibitem[Galli et al.(2016)]{gal16}Galli, A., Vorburger, A., Pommerol, A., Wurz, P., Jost, B., Poch, O., 
Brouet, Y., Tulej, M., Thomas, N., 2016, Surface charging of thick porous water ice layers
relevant for ion sputtering experiments, Planetary and Space Sciences, 126, 63--71.

\bibitem[Galli et al.(2017)]{gal17}Galli, A., Vorburger, A., Wurz, P., Tulej, M., 2017, 
Sputtering of water ice films: A re-assessment with singly and doubly
charged oxygen and argon ions, molecular oxygen, and electrons, Icarus, 291, 36--45.

\bibitem[Hall et al.(1995)]{hal95}Hall, D.T., Strobel, D.F., Feldman, P.D., McGrath, M.A., Weaver, H.A., 1995.
Detection of an oxygen atmosphere on Jupiter's moon Europa. Nature 373, 677.

\bibitem[Hansen et al.(2005)]{han05}Hansen, C.J., Shemansky, D.E., Hendrix, A.R., 2005.
Cassini UVIS observations of Europa's oxygen atmosphere and torus. Icarus 176, 305.

\bibitem[Haring et al.(1984a)]{har84a}Haring, R.A., Kolfschoten, A.W.,
De Vries, A.E., 1984. Chemical sputtering by keV ions. Nuclear Instruments and Methods in Physics Research, B2, 544--549.

\bibitem[Haring et al.(1984b)]{har84b}Haring, R.A., Pedrys, R., Oostra, D.J., Haring, A.,
De Vries, A.E., 1984. Reactive sputtering of simple condensed gases by keV ions III:
kinetic energy distributions. Nuclear Instruments and Methods in Physics Research, B5, 483--488.

%\bibitem[Hollister and Sinano\u{g}lu(1966)]{hol66} Hollister, C. and Sinano\u{g}lu, O., 1966. 
%Molecular binding energies. Journal of the American Chemical Society 88, 1. 

\bibitem[Ip et al.(1998)]{ip98} Ip, W.-H., Williams,  D.J., McEntire, R.W., Mauk, B.H., 1998. Ion
sputtering and surface erosion at Europa. Geophysical Research Letters, 25, 829--832, doi:10.1029/98GL00472.

\bibitem[Johnson(1989)]{joh89}Johnson, R.E., 1989.
Application of Laboratory Data to the Sputtering of a Planetary Regolith. Icarus 78, 206--210.

\bibitem[Johnson et al.(2004)]{joh04}Johnson, R.E., Carlson, R.W., Cooper, J.F., Paranicas, C., Moore, M.H., Wong, M.C., 2004. 
Radiation effects on the surfaces of the Galilean satellites. In: Bagenal, F. (Ed.), 
Jupiter: Atmosphere, Satellites and Magnetosphere. University of Arizona Press, Tucson, USA.

\bibitem[Johnson et al.(2009)]{joh09}Johnson, R.E., Burger, M.H., Cassidy, T.A., 
Leblanc, F., Marconi, M., Smyth, W.H., 2009. Composition 
and Detection of Europa’s Sputter-Induced Atmosphere. In:
Pappalardo, R.T., McKinnon, W.B., Khurana, K.K. (Eds.), Europa. University of
Arizona Press, Tucson.

%\bibitem[K\"ustner et al.(1998)]{kue98} K\"ustner, M., Eckstein, W., Dose, V., Roth, J, 1998.
%The influence of surface roughness on the angular dependence of the sputter yield. Nuclear Instruments and Methods in Physical Research B 145, 320--331.

\bibitem[Marti et al.(2001)]{mar01}Marti, A., Schletti, R., Wurz, P., Bochsler, P., 2001.
Calibration facility for solar wind plasma instrumentation. Review of Scientific Instruments 72, 1354. doi:10.1063/1.1340020.

\bibitem[Mitchell et al.(2017)]{mit17}Mitchell, E.H., Raut, U., Teolis, B.D., Baragiola, R.A., 2017.
Porosity effects on crystallization kinetics of amorphous solid water:
Implications for cold icy objects in the outer solar system. Icarus 285, 291--299.

\bibitem[Muntean et al.(2016)]{mun16}Muntean, E.A., Lacerda, P., Field, T.A.,
 Fitzsimmons, A., Fraser, W.C., Hunniford, A.C., McCullough, R.W., 2016.
A laboratory study of water ice erosion by low-energy ions. Monthly Notices of the Royal Astronomical Society 462, 3361--3367.

\bibitem[Paranicas et al.(2001)]{par01}Paranicas, C., Carlson, R. W., Johnson, R.E., 2001. Electron
bombardment of Europa. Geophysical Research Letters 28, 673--676, doi:10.1029/2000GL012320.

\bibitem[Paranicas et al.(2002)]{par02}Paranicas, C., Mauk, B.H., Ratliff, J.M., Cohen, C., Johnson, R.E., 2002. 
The ion environment near Europa and its role in surface energetics. Geophysical Research Letters 29, 18. 

\bibitem[Plainaki et al.(2012)]{pla12}Plainaki, C., Milillo, A., Mura, A., Orsini, S., Massetti, S., Cassidy, T., 2012.
The role of sputtering and radiolysis in the generation of Europa exosphere. Icarus, 218, 956.

\bibitem[Reimann et al.(1984)]{rei84}Reimann, C.T., Boring, J.W., Johnson, R.E., Garrett, J.W.,
Farmer, K.R., 1984. Ion-induced molecular ejection from D$_{2}$O ice. Surface Science 147, 227--240.

\bibitem[Roth et al.(2016)]{rot16}Roth, L., Saur, J., Retherford, K.D., Strobel, D.F., Feldman, P.D., McGrath, M.A., Spencer, J.R., Blöcker, A.,
Ivchenko, N., 2016. Europa's far ultraviolet oxygen aurora from a comprehensive set of HST observations.
Journal of Geophysical Research 121, 2143--2170.

\bibitem[Shematovich et al.(2005)]{she05}Shematovich, V. I., Johnson, R. E., Cooper, J. F., Wong, M. C., 2005.
Surface-bounded atmosphere of Europa. Icarus 173, 480.

\bibitem[Shi et al.(1995)]{shi95}Shi, M., Baragiola, R. A., Grosjean, D. E., Johnson, R. E., Jurac, S., Schou, J., 1995.
Sputtering of water ice surfaces and the production of extended neutral atmospheres. Journal of Geophysical Research 100, 26,387--26,395.

\bibitem[Smyth and Marconi(2006)]{smy06}Smyth, W.H. and Marconi, M.L., 2006.
Europa's atmosphere, gas tori, and magnetospheric implications. Icarus 181, 510--526.

\bibitem[Sigmund(1969)]{sig69}Sigmund, P. 1969. Theory of sputtering. I. Sputtering yield of amorphous and 
polycrystalline targets. Physical Review 184, 383--416.

\bibitem[Teolis et al.(2005)]{teo05}Teolis, B.D., Vidal, R.A., Shi, J., Baragiola, R.A., 2005.
Mechanisms of O$_{2}$ sputtering from water ice by keV ions. Physical Review B 72, 245422.

\bibitem[Teolis et al.(2009)]{teo09}Teolis, B.D., Shi, J., Baragiola, R.A., 2009. 
Formation, trapping, and ejection of radiolytic O$_{2}$ from ion-irradiated water
ice studied by sputter depth profiling. The Journal of Chemical Physics 130, 134704.

\bibitem[Vidal et al.(2005)]{vid05}Vidal R.A., Teolis, B.D., Baragiola, R.A., 2005.
Angular dependence of the sputtering yield of water ice by 100 keV proton bombardment.
Surface Science 588, 1--5.

\bibitem[Vorburger and Wurz(2018)]{vor18}Vorburger, A., Wurz, P., 2018. Europa's Ice-Related Atmosphere: The Sputter Contribution. 
Icarus 311, 135--145.

\bibitem[Ziegler et al.(1985)]{zie85}Ziegler, J.F., Biersack, J.P., Littmark, U., 1985. 
The stopping and range of ions in matter. Pergamon, New York.

\bibitem[Ziegler et al.(2008)]{zie08}Ziegler, J.F., Biersack, J.P., Ziegler, M. D., 2008.
SRIM -- The Stopping and Range of Ions in Matter, Vol. 5. SRIM Co., Chester, MD.

\end{thebibliography}
\end{document}